\documentclass[12pt,a4paper]{article}
\usepackage[british]{babel}
\usepackage{epsfig}
                                                                                
\begin{document}                                                                
\date{}

\title{ Unquenched domain wall quarks with multi-bosons }

\author{ I.\ Montvay                             \\
         Deutsches Elektronen-Synchrotron DESY   \\
         Notkestr.\,85, D-22603 Hamburg, Germany }

\newcommand{\be}{\begin{equation}}                                              
\newcommand{\ee}{\end{equation}}                                                
\newcommand{\half}{\frac{1}{2}}                                                 
\newcommand{\rar}{\rightarrow}                                                  
\newcommand{\lar}{\leftarrow}
\newcommand{\LCB}{\raisebox{-0.3ex}{\mbox{\LARGE$\left\{\right.$}}}
\newcommand{\RCB}{\raisebox{-0.3ex}{\mbox{\LARGE$\left.\right\}$}}}
\newcommand{\U}{\mathrm{U}}
\newcommand{\SU}{\mathrm{SU}}
                                                                                
\maketitle
\vspace*{1em}

\begin{abstract} \normalsize
 The numerical simulation of domain wall quarks with the two-step
 multi-boson (TSMB) algorithm is considered.
 The inclusion of single quark flavours, as required for strange
 quarks, is discussed.
 The usage of computer memory can be kept relatively low, independently
 of the order of polynomial approximations.
 Tests are performed with two flavours ($N_f=2$) of degenerate quarks
 near the $N_t=4$ thermodynamical cross over.
\end{abstract}       

\section{Introduction}\label{sec1}

 Domain wall fermions \cite{KAPLAN,NARANEU,SHAMIR,FURSHA} offer the
 possibility of improving chiral symmetry of lattice discretizations for
 fermionic theories by tuning the action parameters in an extra (fifth)
 dimension.
 It can be shown \cite{NEUBERGER,KIKUKAWA} that in the limit of
 vanishing lattice spacing in the fifth dimension the domain wall
 formulation is equivalent to the overlap formulation
 \cite{OVERLAP1,OVERLAP2} which fulfills the Ginsparg-Wilson relation
 \cite{GINWIL} for lattice chiral symmetry \cite{LUSCHER}.
 The price of the chiral symmetry at non-zero lattice spacing is the
 extra dimension enlarging the number of degrees of freedom and, from a
 technical point of view, the extensions of the fermion matrix.
 It is an interesting question how much numerical simulations with light
 domain wall quarks become slower than, say, with ``unimproved''
 Wilson fermions.

 Up to now unquenched domain wall fermions have been treated either
 by the Hybrid Monte Carlo \cite{HMC} or, in case of an odd number
 of fermionic flavours, by the Hybrid Molecular Dynamics R-algorithm
 \cite{HMD}.
 (For a few examples of these simulations see, for instance,
 \cite{JASTER,VRANAS,NT4,SYM}.
 For a recent review see \cite{REVIEW}.)
 In the present paper the application of the two-step multi-boson (TSMB)
 algorithm \cite{TSMB} for domain wall fermion simulations is
 considered.
 This algorithm is applicable for any number of fermion flavours and
 has a tolerable slowing down towards light fermions.

 The plan of this paper is as follows: in the next Section the lattice
 action for two degenerate domain wall quarks is formulated.
 In Section~\ref{sec3} the TSMB algorithm is briefly recapitulated and
 the generalization to the case of an odd number of domain wall fermion
 flavours is discussed.
 In Section~\ref{sec4} the results of test runs on $8^3 \cdot 4$
 lattices near the thermodynamical cross over are given.

\section{Lattice action}\label{sec2}

 In this paper the domain wall fermion action is constructed according
 to Shamir's prescription \cite{SHAMIR}.
 Therefore the fermion field is defined on a five dimensional hypercubic
 lattice and the light chiral fermion modes are located on two
 boundaries of the fifth dimension.
 The gauge field links depend only on the coordinates of the
 four-dimensional space-time.
 The bosonic Pauli-Villars fields subtracting the heavy fermion modes
 are introduced as in \cite{VRANAS}.

 The complete lattice action is given by
\be  \label{eq01}
S = S_G[U] + S_F[\overline{\Psi},\Psi,U] 
+ S_{PV}[\Phi^\dagger,\Phi,U] \ .
\ee
 Here the standard Wilson action $S_G$ for the SU($N_c$) gauge field
 $U$ is a sum over plaquettes
\be  \label{eq02}
S_G  =   \beta \sum_{pl}                                                  
\left( 1 - \frac{1}{N_c} {\rm Re\,Tr\,} U_{pl} \right) \ ,   
\ee
 with the bare gauge coupling given by $\beta \equiv 2N_c/g^2$.
 In particular, in QCD the number of colours is $N_c=3$.
 The Pauli-Villars action in (\ref{eq01}) $S_{PV}$ will be discussed
 later and the fermion action $S_F$ for a single fermion flavour is
 given by
\be  \label{eq03}
S_F = \sum_{x,s;\,x^\prime,s^\prime} \overline{\Psi}(x^\prime,s^\prime)
D_F(x^\prime,s^\prime;\,x,s) \Psi(x,s) \ .
\ee
 The four-dimensional space-time coordinates are denoted by $x,x^\prime$
 and the fifth coordinates are $s,s^\prime$.
 The domain wall fermion matrix $D_F$ is constructed from the standard
 four-dimensional Wilson fermion matrix
\be  \label{eq04}
D(x^\prime,x) =
\delta_{x^\prime x}\; (4-am_0) - \half \sum_{\mu=1}^4 \left[
\delta_{x^\prime,x+\hat{\mu}}(1+\gamma_\mu) U_{x\mu} +
\delta_{x^\prime+\hat{\mu},x}(1-\gamma_\mu) U^\dagger_{x^\prime\mu}
\right] \ .
\ee
 The notations are standard: $a$ is the (four-dimensional) lattice
 spacing and $\hat{\mu}$ denotes the unit vector in direction $\mu$.
 The bare mass $-m_0$ is chosen to be negative and should be tuned
 properly for producing the light boundary fermion state.
 In an $s$-block notation the domain wall fermion matrix is
\be  \label{eq05}
D_F = \left( \begin{array}{cccccccc}
\sigma+D    & -\sigma P_L & 0           & 0           & \ldots & 
0           & 0           &  am_f P_R   
   \\[0.7em]
-\sigma P_R & \sigma+D    & -\sigma P_L & 0           & \ldots & 
0           & 0           &  0          
   \\[0.7em]
0           & -\sigma P_R & \sigma+D    & -\sigma P_L & \ldots & 
0           & 0           &  0          
   \\[0.7em]
\vdots      & \vdots      & \vdots      & \vdots      & \ddots & 
\vdots      & \vdots      & \vdots      
   \\[0.7em]
0           & 0           & 0           & 0           & \ldots & 
-\sigma P_R & \sigma+D    & -\sigma P_L 
   \\[0.7em]
am_f P_L    & 0           & 0           & 0           & \ldots & 
0           & -\sigma P_R & \sigma+D
\end{array} \right) \ .
\ee
 Here $m_f$ denotes the bare fermion mass of the light boundary fermion,
 $P_R = \half(1+\gamma_5),\; P_L = \half(1-\gamma_5)$ are chiral
 projectors and $\sigma \equiv a/a_s$ determines the lattice spacing in
 the fifth dimension $a_s$ relative to $a$ \cite{KIKYAM}.

 The domain wall fermion matrix $D_F$ is non-hermitean but it satisfies
 the relation
\be  \label{eq06}
D_F^\dagger = \gamma_5 R_5\, D_F\, \gamma_5 R_5
\ee
 with the reflection in the fifth dimension
 $(R_5)_{s^\prime,s} \equiv \delta_{s^\prime,N_s+1-s},\;
 (1 \leq s \leq N_s)$.
 This relation implies that the determinant of $D_F$ is real and
 $\tilde{D}_F \equiv \gamma_5 R_5\, D_F$ is hermitean.
 Using the hermitean Wilson fermion matrix
 $\tilde{D} \equiv \gamma_5 D$ the hermiticity of $\tilde{D}_F$
 in nicely displayed in an $s$-block form:
\be  \label{eq07}
\tilde{D}_F = \left( \begin{array}{cccccccc}
-am_f P_L                & 0                        & 0          &
\ldots & 0                        & 0                        &
-\sigma P_R              & \sigma\gamma_5+\tilde{D} \\[0.7em]
0                        & 0                        & 0          &
\ldots & 0                        & -\sigma P_R              &
\sigma\gamma_5+\tilde{D} & \sigma P_L               \\[0.7em]
0                        & 0                        & 0          &
\ldots & -\sigma P_R              & \sigma\gamma_5+\tilde{D} &
\sigma P_L               & 0                        \\[0.7em]
0                        & 0                        & 0          &
\ldots & \sigma\gamma_5+\tilde{D} & \sigma P_L               &
0                        & 0                        \\[0.7em]
\vdots                   & \vdots                   & \vdots     &
\ddots & \vdots                   & \vdots                   &
\vdots                   & \vdots                   \\[0.7em]
-\sigma P_R              & \sigma\gamma_5+\tilde{D} & \sigma P_L &
\ldots & 0                        & 0                        &
0                        & 0                        \\[0.7em]
\sigma\gamma_5+\tilde{D} & \sigma P_L               & 0          &
\ldots & 0                        & 0                        &
0                        & am_f P_R                 
\end{array} \right) \ .
\ee

 The Pauli-Villars action is designed to cancel the contribution of the
 heavy fermions in the large $N_s$ limit.
 It has a Gaussian form resulting after integration in the inverse of
 the fermion determinant.
 In order that the Gaussian integrals be well defined let us first
 only consider the case of two degenerate fermion flavours when
 the Pauli-Villars action is
\be  \label{eq08}
S_{PV} = \sum_{x,s;\,x^\prime,s^\prime;\,
x^{\prime\prime},s^{\prime\prime}}
\Phi^\dagger(x^{\prime\prime},s^{\prime\prime})\;
D_F^\dagger(x^{\prime\prime},s^{\prime\prime};x^\prime,s^\prime)_{am_f=1}
\; D_F(x^\prime,s^\prime;\,x,s)_{am_f=1}\; \Phi(x,s) \ .
\ee
 As it is shown by the notation, the bare mass parameter $am_f$ is
 fixed here at $am_f=1$.
 After performing the Grassmannian integrals over the fermion fields
 $\overline{\Psi},\Psi$ and the Gaussian integrals over the
 Pauli-Villars fields $\Phi^\dagger,\Phi$ the result, for two degenerate
 fermion flavours, is the ratio of determinants
\be  \label{eq09}
\frac{\det(D_F)\det(D_F)}{\det(D^\dagger_{F,am_f=1} D_{F,am_f=1})} =
\frac{\det(\tilde{D}_F^2)}{\det(\tilde{D}_{F,am_f=1}^2)} \ .
\ee
 Here we used the fact that $\det(D_F)$ is real and
 $\det(D_F)=\det(\tilde{D}_F)$.

 According to (\ref{eq09}) the effective gauge action describing two
 degenerate flavours of fermions is a function of the squared hermitean
 fermion matrix $\tilde{D}_F^2$.
 The same is also true in case of any number of flavours, as it will
 be discussed in the next section.
 The non-zero matrix elements of $\tilde{D}_F^2$ are, in an $s$-block
 form:
\begin{eqnarray}  \label{eq10}
(\tilde{D}_F^2)_{s,s} &=& 2\sigma^2+2\sigma D_r+\tilde{D}^2
+\delta_{s,1}  (a^2m_f^2-\sigma^2)P_L
+\delta_{s,N_s}(a^2m_f^2-\sigma^2)P_R \ ,
\nonumber \\[0.7em]
(\tilde{D}_F^2)_{s,s+1} &=& (\tilde{D}_F^2)_{s+1,s} =
-\sigma^2-\sigma D_r \ ,
\nonumber \\[0.7em]
(\tilde{D}_F^2)_{1,N_s} &=& (\tilde{D}_F^2)_{N_s,1} =
am_f(\sigma+D_r) \ ,
\end{eqnarray}
 where $D_r= \half(\gamma_5\tilde{D}+\tilde{D}\gamma_5)$ contains the
 Wilson term in the Wilson-fermion action:
\be  \label{eq11}
D_r(x^\prime,x) =
\delta_{x^\prime x}\; (4-am_0) - \half \sum_{\mu=1}^4 \left[
\delta_{x^\prime,x+\hat{\mu}} U_{x\mu} +
\delta_{x^\prime+\hat{\mu},x} U^\dagger_{x^\prime\mu}
\right] \ .
\ee
%

\section{TSMB algorithm for domain wall fermions}\label{sec3}

 The absolute value of the fermion determinant of an arbitrary (integer)
 number $N_f$ of domain wall fermion flavours is, according to
 (\ref{eq09}),
\be  \label{eq12}
\left|\det(\tilde{D}_F)\right|^{N_f} =
\left\{\det(\tilde{D}_F^2)\right\}^{N_f/2} \ .
\ee
 Negative values of $N_f$ describe the Pauli-Villars fields.
 (The mass parameter $am_f$ is different for physical fermion flavours
 and for Pauli-Villars fields, but this difference does not play a
 r\^ole in what follows.)
 For odd numbers of flavours the sign of the determinant is neglected
 in (\ref{eq12}).
 However, if the mass parameter is positive ($m_f>0$) this sign is
 expected to be irrelevant because of the relation of domain wall
 fermions to overlap fermions which have a positive determinant if the
 mass is positive \cite{NEUBERGER,KIKUKAWA}.

 Multi-boson algorithms \cite{MULTIB} are based on polynomial
 approximations $P_n$ satisfying
\be  \label{eq13}
\lim_{n \to \infty} P_n(x) = x^{-N_f/2}
\ee
 which allow to represent the fermion determinant as
\be  \label{eq14}
\left\{\det(\tilde{D}_F^2)\right\}^{N_f/2}
\simeq \frac{1}{\det P_n(\tilde{D}_F^2)} \ .
\ee
 Assuming that the polynomial roots occur in complex conjugate pairs,
 one can write $P_n$ as
\be \label{eq15}
P_n(\tilde{D}_F^2) \propto
\prod_{j=1}^n (\tilde{D}_F-\rho_j^*) (\tilde{D}_F-\rho_j) \ .
\ee
 This leads to the {\em multi-boson representation}
\begin{eqnarray} \label{eq16}
\left|\det(\tilde{D}_F)\right|^{N_f} &\propto&
\prod_{j=1}^n\det[(\tilde{D}_F-\rho_j^*) (\tilde{D}_F-\rho_j)]^{-1} 
\nonumber \\[0.5em]
&\propto&
\int [d\phi][d\phi^+]\; \exp\left\{ -\sum_{j=1}^n \sum_{x x^\prime}
\phi_{j x^\prime}^+\, [(\tilde{D}_F-\rho_j^*) 
(\tilde{D}_F-\rho_j)]_{x^\prime x}\, \phi_{jx} \right\} \ .
\end{eqnarray}
 Here $\phi_{jx},\;(j=1,2,\ldots,n)$ are complex boson (pseudofermion)
 fields.

 The two-step multi-boson algorithm \cite{TSMB} is based, instead of
 (\ref{eq13}), on a polynomial approximation by a product of
 polynomials
\be \label{eq17}
\lim_{n_2 \to \infty} P^{(1)}_{n_1}(x)P^{(2)}_{n_2}(x) =
x^{-N_f/2} \ ,
\ee
 where the first polynomial $P^{(1)}_{n_1}(x)$ itself is an
 approximation to $x^{-N_f/2}$, but it has a relatively low order.
 The multi-boson representation (\ref{eq16}) is only used for the first
 polynomial $P^{(1)}_{n_1}$.
 The correction factor $P^{(2)}_{n_2}$ in
\be \label{eq18}
\left|\det(\tilde{D}_F)\right|^{N_f} \;\simeq\;
\frac{1}{\det P^{(1)}_{n_1}(\tilde{D}_F^2)\;
\det P^{(2)}_{n_2}(\tilde{D}_F^2)}
\ee
 is realized in a stochastic {\em noisy Metropolis correction step} with
 a global accept-reject condition, in the spirit of \cite{CORR}.
 In order to obtain the appropriate Gaussian vector for the noisy
 correction the inverse square root of $P^{(2)}_{n_2}$ is also needed.
 This can be represented by another polynomial approximation
\be \label{eq19}
 P^{(3)}_{n_3}(x) \simeq P^{(2)}_{n_2}(x)^{-\half} \ .
\ee
 A practical way to obtain $P^{(3)}$ is to use a Newton iteration
\be \label{eq20}
P^{(3)}_{k+1} = \half \left( P^{(3)}_k + \frac{1}{P^{(3)}_k P^{(2)}} 
\right) \ , \hspace{3em} k=0,1,2,\ldots  \ . 
\ee

 The TSMB algorithm becomes exact only in the limit of infinitely high
 polynomial order: $n_2\to\infty$ in (\ref{eq17})-(\ref{eq18}).
 Instead of investigating the dependence of expectation values on $n_2$
 by performing several simulations, it is better to fix some relatively
 high order $n_2$ for the simulation and perform another correction in
 the ``measurement'' of expectation values by still finer polynomials.
 This is done by {\em reweighting} the configurations \cite{PHMC}.
 The reweighting for general $N_f$ is based on a polynomial
 approximation $P^{(4)}_{n_4}$ which satisfies
\be\label{eq21}
\lim_{n_4 \to \infty} P^{(1)}_{n_1}(x)P^{(2)}_{n_2}(x)P^{(4)}_{n_4}(x) =
x^{-N_f/2} \ .
\ee
 For more details see, for instance, \cite{POLYNOM,VARYACT}.

 Up to this point the particular form of the domain wall fermion
 action introduced in the previous section makes no difference compared
 to other applications of TSMB.
 The occurrence of negative number of flavours, as used for the
 Pauli-Villars fields, is the only new feature.
 However, for even number of flavours one can just use the form given
 by (\ref{eq09}).
 In this case the first polynomial $P^{(1)}$ for the Pauli-Villars
 fields is exact and no corrections are needed.
 For odd number of flavours one has to deal with polynomial
 approximations of some integer power of $\sqrt{x}$ and the machinery
 of polynomial approximations works as usual.

 Another peculiarity of domain wall fermions is that the fermion field
 has an extra index labeling the fifth coordinate.
 In practice this can easily lead to a situation where the storage
 of $n_1$ multi-boson fields in computer memory becomes a problem.
 (For typical simulation parameters including the values of $n_1$
 see the recent studies in \cite{NF3,QED,SCHROERS}.)
 Fortunately, in cases if the storage of the multi-boson fields is
 problematic, one can organize the gauge field update in such a way
 that the dependence on the multi-boson fields is collected in a few
 auxiliary $3 \otimes 3$ matrix fields which can be easily stored in
 memory.
 The multi-boson fields can be kept on disk and have to be read before
 and written back after a complete boson field update.
 The duration of the input-output is negligible compared to the time
 of the update.

 The auxiliary $3 \otimes 3$ matrix fields are spin-traces over the
 multi-boson fields which can be constructed as follows.
 The dependence of the effective gauge field action on the multi-boson
 fields can be summarized by the formula
\be\label{eq22}
S_{eff}(U_{x\mu},\phi) = {\rm Re\, Tr\,} \left(
U_{x\mu} S^{(1)}_{x\mu}(\phi) + U_{x\mu} S^{(2)}_{x\mu}(\phi) \right)
+ \cdots \ .
\ee
 Here $U_{x\mu}$ is the gauge link variable to be updated.
 The omitted part denoted by the dots contain terms from the pure gauge
 action and other terms which do not depend on $U_{x\mu}$.
 $S^{(1,2)}_{x\mu}(\phi)$ display the dependence on the multi-boson
 fields.
 $S^{(1)}$ gives the nearest neighbour contributions and $S^{(2)}$,
 which appears because of the quadratic nature of $\tilde{D}_F^2$,
 stands for next--to--nearest neighbour contributions.
 A consideration of the multi-boson action (\ref{eq16}) and of the
 matrices in (\ref{eq07}), (\ref{eq10}) gives that
\be\label{eq23}
S^{(1)}_{x\mu}(\phi) = \sum_{s=1}^{N_s} \left(
\tilde{f}_{s,N_s+1-s;x\mu} + \chi f^{(r)}_{s,s;x\mu} \right)
+ \mu_f f^{(r)}_{1,N_s;x\mu}
- \sigma \sum_{s=1}^{N_s-1} f^{(r)}_{s,s+1;x\mu}
\ee
 and
\be\label{eq24}
S^{(2)}_{x\mu}(\phi) = \sum_{\nu\ne\mu,\, \nu=1}^4 \left(
f^{(++)}_{x,\mu\nu}\; U_{x+\hat{\mu},\nu} +
f^{(+-)}_{x,\mu\nu}\; U^\dagger_{x+\hat{\mu}-\hat{\nu},\nu} +
U^\dagger_{x,\nu}\; f^{(-+)}_{x+\hat{\nu},\nu\mu} +
U_{x-\hat{\nu},\nu}\; f^{(++)}_{x-\hat{\nu},\nu\mu} \right) \ .
\ee
 Here we used the notations
\be\label{eq25}
\mu_f \equiv am_f \ , \hspace{2em} \mu_0 \equiv am_0 \ , \hspace{2em}
\chi \equiv 4-\mu_0+\sigma \ . 
\ee
 The traces over spinor indices appearing in (\ref{eq23}), resp.
 (\ref{eq24}) are defined as
\begin{eqnarray} \label{eq26}
\tilde{f}_{s^\prime,s;x\mu} &\equiv& - \sum_{j=1}^{n_1} 
{\rm Re\,}\rho_j\; {\rm Tr_{sp}\,}\left[ (\gamma_5+\gamma_5\gamma_\mu)\;
\phi_{js,x}\; \phi^\dagger_{js^\prime,x+\hat{\mu}} \right]
+ (s \leftrightarrow s^\prime) \ ,
\nonumber \\[0.5em]
f^{(r)}_{s^\prime,s;x\mu} &\equiv& - \sum_{j=1}^{n_1} {\rm Tr_{sp}\,}
\left[ \phi_{js,x}\; \phi^\dagger_{js^\prime,x+\hat{\mu}} \right]
+ (s \leftrightarrow s^\prime) \ ,
\nonumber \\[0.5em]
f^{(++)}_{x,\mu\nu} &\equiv& \half\sum_{j=1}^{n_1}\sum_{s=1}^{N_s}
{\rm Tr_{sp}\,}\left[ (1+\gamma_\mu-\gamma_\nu-\gamma_\nu\gamma_\mu)\;
\phi_{js,x}\; \phi^\dagger_{js,x+\hat{\mu}+\hat{\nu}} \right] \ ,
\nonumber \\[0.5em]
f^{(+-)}_{x,\mu\nu} &\equiv& \half\sum_{j=1}^{n_1}\sum_{s=1}^{N_s}
{\rm Tr_{sp}\,}\left[ (1+\gamma_\mu+\gamma_\nu+\gamma_\nu\gamma_\mu)\;
\phi_{js,x}\; \phi^\dagger_{js,x+\hat{\mu}-\hat{\nu}} \right] \ ,
\nonumber \\[0.5em]
f^{(-+)}_{x,\nu\mu} &\equiv& \half\sum_{j=1}^{n_1}\sum_{s=1}^{N_s}
{\rm Tr_{sp}\,}\left[ (1-\gamma_\nu-\gamma_\mu+\gamma_\nu\gamma_\mu)\;
\phi_{js,x}\; \phi^\dagger_{js,x-\hat{\nu}+\hat{\mu}} \right] \ .
\end{eqnarray}
 The indices on the multi-boson fields $\phi_{js,x}$ are as follows:
 $j$ is labeling the different multi-boson fields as they appear in
 (\ref{eq16}), $x$ is the four-dimensional site and $s$ the fifth
 coordinate.
 The colour and spinor indices are not shown.
 After performing the trace over spinor indices the result is, of
 course, a $3 \otimes 3$ complex matrix.

 Using the formulas (\ref{eq23})-(\ref{eq26}) the effect of the
 multi-boson fields can be collected in the auxiliary $3 \otimes 3$
 matrices.
 The total number of $3 \otimes 3$ matrices to be stored in memory is
 $4+3 \cdot 12=40$ per four-dimensional site, which is usually not a
 problem.
 The multi-boson fields can be updated one-by-one and kept otherwise
 on disk.
 This means that one has to store $\phi_{js,x}$ only for a single value
 of the multi-boson index $j$.
 Of course, due to the index $s$, the storage requirement is increasing
 with $N_s$.

\section{Numerical simulation tests}\label{sec4}

 Test runs have been performed for two degenerate quark flavours $N_f=2$
 on $8^3 \cdot 4$ lattices in the vicinity of the $N_t=4$ thermodynamic
 crossover.
 The parameter sets have been chosen from the points in parameter space
 which were investigated in \cite{NT4}.
 Typical parameters were: $\mu_0=1.9,\;\mu_f=0.1,\;\sigma=1.0,\,0.5$ and
 $5.20 \leq \beta \leq 5.45$.

 The first task is to determine the parameters of the necessary
 polynomials for different extensions of the fifth dimension $N_s$.
 These will be largely influenced by the condition number of the squared
 hermitean fermion matrix $\tilde{D}_F^2$, because the interval
 $[\epsilon,\lambda]$ where the polynomial approximations are optimized
 has to cover the eigenvalues of $\tilde{D}_F^2$ on a typical gauge
 configuration.
 A set of polynomial parameters for different $N_s$ is given in
 table~\ref{tab01}.
 As it is shown by the table, the largest eigenvalues of $\tilde{D}_F^2$
 are only slightly increasing with $N_s$ but, in the covered range,
 the smallest ones decrease roughly proportional to $1/N_s$.
 Therefore the condition number $\lambda/\epsilon$ is proportional
 to $N_s$.
 (For a discussion of bounds on the condition number see
 \cite{CONDITION}.)
\begin{table}[ht]
\begin{center}
\parbox{15cm}{\caption{\label{tab01}\em
 Parameters of the polynomials used on $8^3 \cdot 4$ lattice for
 different lattice spacings and extensions of the fifth dimension.
 The bare parameters in the lattice action are specified in the text.
 The last column shows the average number of inverter iterations in the
 quasi heatbath update step.
}}
\end{center}
\begin{center}
\begin{tabular}{| l | r || r | r | r | r | l | l || r |}
\hline
\multicolumn{1}{|c|}{$\sigma$}   & \multicolumn{1}{|c||}{$N_s$} &
\multicolumn{1}{|c|}{$n_1$}      & \multicolumn{1}{|c|}{$n_2$}  &
\multicolumn{1}{|c|}{$n_3$}      & \multicolumn{1}{|c|}{$n_4$}  &
\multicolumn{1}{|c|}{$\epsilon$} & \multicolumn{1}{|c|}{$\lambda$} &
\multicolumn{1}{||c||}{$I_{QHB}$} \\
\hline\hline
1.0 & 8  & 44  &  240  &  300  &  300  &  0.011   &  56.0  & 5000  \\
\hline
1.0 & 12 & 56  &  300  &  450  &  470  &  0.0071  &  57.0  & 7900  \\
\hline
1.0 & 16 & 64  &  350  &  470  &  500  &  0.0052  &  58.0  & 9700  \\
\hline
1.0 & 24 & 72  &  450  &  640  &  640  &  0.0032  &  59.0  & 12300 \\
\hline\hline
0.5 & 6  & 48  &  270  &  350  &  360  &  0.0071  &  43.0  & 6000  \\
\hline
0.5 & 12 & 64  &  400  &  570  &  570  &  0.0046  &  44.0  & 10100 \\
\hline\hline
\end{tabular}
\end{center}
\end{table}

 Table~\ref{tab01} refers to the case $N_f=2$ which is considered in the
 present paper.
 As discussed in the previous section, an odd number of flavours would
 also require polynomial approximations for the Pauli-Villars fields.
 Since the mass parameter is kept there at $\mu_f=1$, the corresponding
 condition numbers are much smaller and the polynomial orders
 substantially lower.
 In the runs shown by table~\ref{tab01} the ratio of condition numbers
 is typically of the order of 10.
 This would require, for instance, in case of the first line of the
 table polynomial orders $n_1=32,\; n_2=60,\;n_3=90,\;n_4=100$.

 A first estimate of the computation work can be given in terms of the
 number of fermion-matrix vector multiplications needed for performing a
 sweep over the multi-boson and gauge fields.
 An approximate formula per update cycles is:  
\be\label{eq27}
N_{MVM} \simeq 2I_{QHB}c_{QHB}^{-1} + 
6(n_1 N_B + N_G) + 2N_C (n_2+n_3) \ .
\ee
 Here $N_B$ is the number of boson field update sweeps per cycle, $N_G$
 the number of gauge field update sweeps and $N_C$ the number of noisy
 Metropolis accept-reject steps.
 It is assumed that a global quasi heatbath for the boson fields
 \cite{QUASIHTB} is performed once per $c_{QHB}$ cycles with a average
 number of matrix inverter iterations $I_{QHB}$.
 (Note that $I_{QHB}$ is a sum over $n_1$ inversions.)
 The relative contributions of the three terms in (\ref{eq27}) are
 subject to optimization.
 The quasi heatbath is typically an important part of operations and
 $I_{QHB}$ is characteristic for the total number of matrix vector
 multiplications $N_{MVM}$.
 For the actual values occurring in the runs see table~\ref{tab01}
 which shows that, for a given value of lattice spacing ratio $\sigma$,
 $I_{QHB}$ increases somewhat faster than $N_s^\half$.

 In order to obtain an estimate for the total amount of necessary
 computation work, $N_{MVM}$ has to be multiplied by the integrated
 autocorrelation length $\tau_{int}$ given in number of update cycles.
 The measurement of autocorrelations requires high statistics and a lot
 of computer time and is beyond the scope of the present paper.
 A rough estimate based on previous experience with TSMB tells that
 $\tau_{int}$ is proportional to $n_1$.
 According to the table a very rough estimate for the increase of $n_1$
 is $n_1 \propto N_s^\half$.
 Taking into account that a single matrix vector multiplication grows
 linearly with $N_s$, this would finally imply that the increase
 of the necessary computation work is between $\propto N_s^2$ and
 $\propto N_s^3$.
 The fast increase with $N_s$ favors domain wall fermion schemes with
 $\sigma<1$ and relatively small $N_s$, as proposed in \cite{BETTER}.

 As these test runs show, the application of the TSMB algorithm for
 numerical simulations of domain wall quarks is possible.
 Simulations using alternative formulations based on overlap fermions,
 as proposed in \cite{MASSLESS,CONDITION}, should also be feasible along
 the same lines.
 The interesting question about the computation speed compared to,
 say, Wilson quarks requires a more detailed analysis including
 the measurement of autocorrelations.
 Since the good chiral properties of domain wall fermions develop
 only sufficiently close to the continuum limit, the performance
 studies have to be finally performed on large lattices.

\vspace*{1em}\noindent
{\large\bf Acknowledgement}

\noindent
 I thank Yigal Shamir for helpful discussions and correspondence.




\end{document}